\documentclass[aps,amsmath,amssymb,prl, showpacs, twocolumn,superscriptaddress]{revtex4}

\usepackage{amsfonts}
\usepackage{amsmath}
\usepackage{graphicx}
\usepackage[usenames]{color}

\newcommand{\rvec}{\mathbf{r}}

\newcommand{\dvec}{\mathbf{d}}

\begin{document}

\title{Competition between d-wave and topological p-wave superconductivity in the doped Kitaev-Heisenberg model}

\author{Timo Hyart}
\affiliation{Institut f\"ur Theoretische Physik, Universit\"at Leipzig, D-04103, Leipzig, Germany}
\author{Anthony R. Wright}
\affiliation{Institut f\"ur Theoretische Physik, Universit\"at Leipzig, D-04103, Leipzig, Germany}
\author{Giniyat Khaliullin}
\affiliation{Max-Planck-Institut f\"ur Festk\"orperforschung, D-70569 Stuttgart, Germany}
\author{Bernd Rosenow}
\affiliation{Institut f\"ur Theoretische Physik, Universit\"at Leipzig, D-04103, Leipzig, Germany}

\date{October 4, 2011}

\pacs{74.20.Rp, 74.25.Dw, 75.10.Jm, 72.25.-b}

\begin{abstract}
The competition between Kitaev and Heisenberg interactions away from half filling is studied for the hole-doped Kitaev-Heisenberg $t$-$J_K$-$J_H$ model on a honeycomb lattice. While the isotropic Heisenberg coupling supports a time-reversal violating d-wave singlet state, we find that the Kitaev interaction favors a time-reversal invariant p-wave superconducting phase,  which obeys the rotational symmetries of the microscopic model, and is robust for $J_H<J_K/2$. Within the p-wave superconducting phase, a critical chemical potential  $|\mu|=\mu_c \approx t$ separates a topologically trivial phase for $|\mu|< \mu_c$ from a topologically non-trivial $Z_2$ time-reversal invariant spin-triplet phase for $|\mu|>\mu_c$. 

\end{abstract}

\maketitle

The concept of topological order, originally introduced in the context of quantum Hall systems, has recently been 
extended to topological band insulators and topological superconductors \cite{HaKa10,Zhang-review}. One of the exciting properties of these systems is the prospect of observing non-abelian quasi-particles. A prototype model for non-abelian quasi-particles is the Kitaev model on the honeycomb lattice with a non-abelian topological phase \cite{Kitaev06}.

There is both experimental \cite{SiGe10,Singh+11}  and theoretical \cite{JaKh09} evidence that a
Kitaev model with an admixture of anti-ferromagnetic Heisenberg
exchange is realized in iridates (Li$_2$IrO$_3$ and Na$_2$IrO$_3$) with a
honeycomb lattice. For dominant Kitaev coupling, the ground state is a
spin liquid. Upon increasing the relative strength of the
nearest-neighbor Heisenberg coupling, the ground state first turns
into a stripy antiferromagnetic phase and then into a Neel
antiferromagnet  \cite{Cha+10,Reu+11}.  Both Li$_2$IrO$_3$ and Na$_2$IrO$_3$ are found to be
magnetically ordered at low temperatures  \cite{SiGe10,Singh+11, Liu+11, Ye12, Choi12}. The temperature
dependence of the magnetic susceptibility \cite{SiGe10,Singh+11}  and the recent
resonant x-ray and neutron scattering experiments \cite{Liu+11, Ye12, Choi12} indicate
that the ground state of Na$_2$IrO$_3$ is most likely described by a zig-zag
spin structure. This type of magnetic order can be theoretically
explained by including further neighbor Heisenberg couplings  \cite{Singh+11, Kimchi11}
or in terms of trigonal distortion of the oxygen octahedra  \cite{Bhatta11}.  On the
other hand, based on the magnetic susceptibility measurements for
Li$_2$IrO$_3$, it has been suggested that Li$_2$IrO$_3$ may be close to the Kitaev
spin liquid regime \cite{Singh+11}.

\begin{figure}[tbp]
\centering\includegraphics[width=7cm]{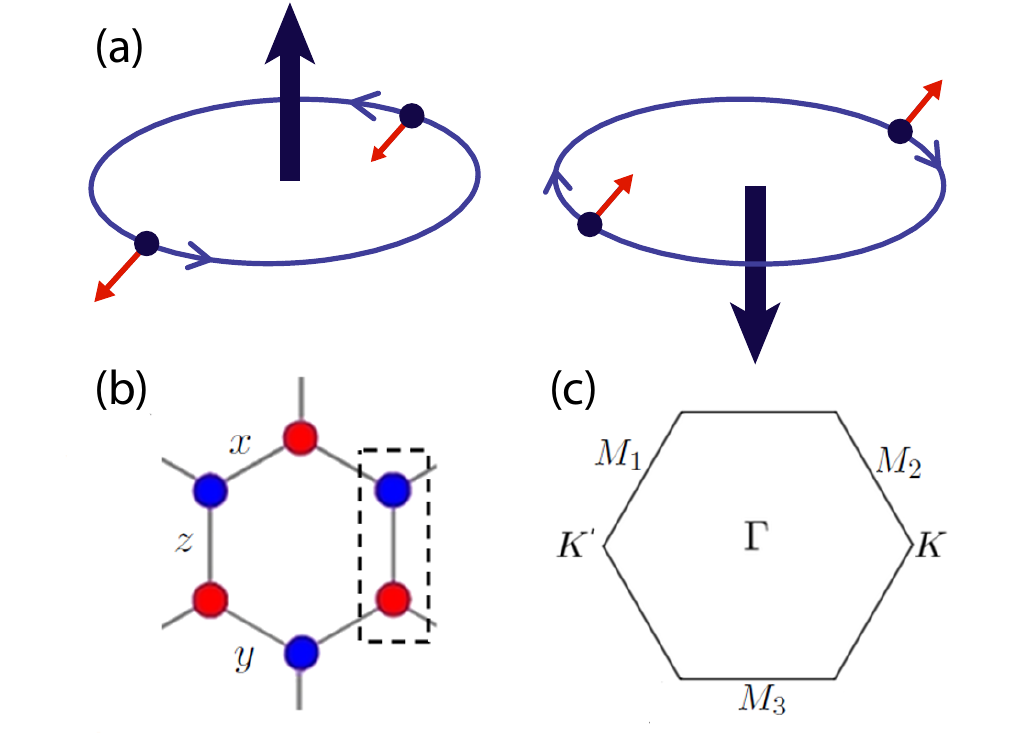}
\caption{(color online) (a) Relative orientations of the Cooper pair orbital (thick arrows) and spin (thin red 
arrows) angular momentum  in the ground state wave function 
for topologically nontrivial time-reversal invariant $p$-wave pairing. 
Conventions for (b) real space and (c) reciprocal lattices. The four time 
reversal invariant points in the Brillouin zone are $\Gamma$, $M_1$, $M_2$, and 
$M_3$. }
\label{pairing.fig}
\end{figure}

The Kitaev spin model can be viewed as a Mott insulator at half filling, and
one can ask about its phase diagram away from half filling.
It is known that a Heisenberg antiferromagnet on the honeycomb lattice
becomes a d-wave superconductor upon doping \cite{BlSc07}. Hence one may
expect superconductivity in the  doped Kitaev model, and can ask how the
superconducting order parameter depends on the relative strengths of
antiferromagnetic Heisenberg exchange and the Kitaev interaction. 
In this paper, we address this question by  
studying the Heisenberg-Kitaev model in a slave boson theory, and find that
the Kitaev model turns into a p-wave superconductor upon doping, with an
internal spin-orbital structure of Cooper pairs as illustrated in  
Fig.~\ref{pairing.fig}(a). When the 
doping level is larger than one quarter, this superconductor is an example of a
time-reversal invariant $Z_2$ topological superconductor \cite{Schny+08,Roy08, Qi2009,
 Sato09a, Sato09, Qi2010, YTada, Sato10}, which supports   
a pair of counter-propagating helical Majorana modes at the edges of the sample
and one pair of Majorana zero modes at the vortices.
This p-wave superconductor is considerably more robust to adding a Heisenberg
exchange than the spin liquid phase itself at zero doping, thus raising the
possibility that it might be observable in doped iridates.  We also find that the time-reversal invariant topological superconductivity is not sensitive to the details of the band structure. 
Thus, the physics discussed here should be representative for a broad class of materials with strong spin-orbit coupling.


We consider a honeycomb lattice with three inequivalent nearest-neighbor bonds referred to as $\gamma = x,y,z$ [see Fig.~\ref{pairing.fig}(b)]. 
The Kitaev-Heisenberg model  with nearest neighbor hopping is given by
\begin{equation}
H = -J_K \sum_{\langle ij\rangle} S_i^\gamma S_j^\gamma + J_H \sum_{\langle ij\rangle} \bigg(\vec{S}_i\cdot \vec{S}_j -\frac{n_i n_j}{4} \bigg) + H_T.
\label{hkh}
\end{equation}
The first term in the Hamiltonian (\ref{hkh}) is called the Kitaev interaction
\cite{Kitaev06}, and it describes an Ising-like  coupling between the $\gamma$
components of spins  $S_i^\gamma = \frac{1}{2} f^\dag_{i,
 \alpha}\sigma^\gamma_{\alpha\beta}f_{i,\beta}$  at each bond in the 
$\gamma$-direction. The second term describes an isotropic Heisenberg interaction
with interaction strength $J_H$, and $H_T$ is the kinetic term  
\begin{equation}
H_T=-t \sum_{<i j>, \sigma} f_{i, \sigma}^\dagger f_{j, \sigma} +h.c. \ .
\end{equation}
We assume that at half-filling the  system can be viewed as a Mott insulator 
and adopt the so-called U(1) slave boson method to take into account the
renormalization of the hopping amplitude by strong correlations 
\cite{bask, kotliar}. By assuming that the holons are condensed to give 
a coherent Fermi-liquid state, the hopping amplitude $t$ can be written 
as $t=t_0 \delta$, where $t_0$ is the bare matrix 
element and $\delta$ quantifies the hole doping so that $1-\delta$ is 
the average number of electrons per site.     

We introduce a spin-singlet and three spin-triplet operators, defined
respectively as $s_{ij} = (f_{i\uparrow}f_{j\downarrow} -
f_{i\downarrow}f_{j\uparrow})/\sqrt{2}$, $t_{ij,x}
=(f_{i\downarrow}f_{j\downarrow} - f_{i\uparrow}f_{j\uparrow}) /\sqrt{2}$,
$t_{ij,y} = i(f_{i\downarrow}f_{j\downarrow} +
f_{i\uparrow}f_{j\uparrow})/\sqrt{2}$, $t_{ij,z} =
i(f_{i\uparrow}f_{j\downarrow} + f_{i\downarrow}f_{j\uparrow})/\sqrt{2}$, and
re-write the Kitaev-Heisenberg part of the Hamiltonian in terms of these
operators.  The Kitaev term on an $x$-link is given by 

\begin{equation}
-S_i^xS_j^x = \frac{1}{4}\bigl[s_{ij}^\dag s_{ij} + t_{ij,x}^\dag t_{ij,x} -
t_{ij,y}^\dag t_{ij,y} - t_{ij,z}^\dag t_{ij,z}  \bigr]\  ,   
\end{equation}
and for $ y$- and $z$-links the plus sign appears in front of the
$t_{ij,y}^\dagger t_{ij,y}$ and $t^\dagger_{ij,z}t_{ij,z}$ operators, 
respectively.

\begin{figure}[tbp]
\centering\includegraphics[width=6.5cm]{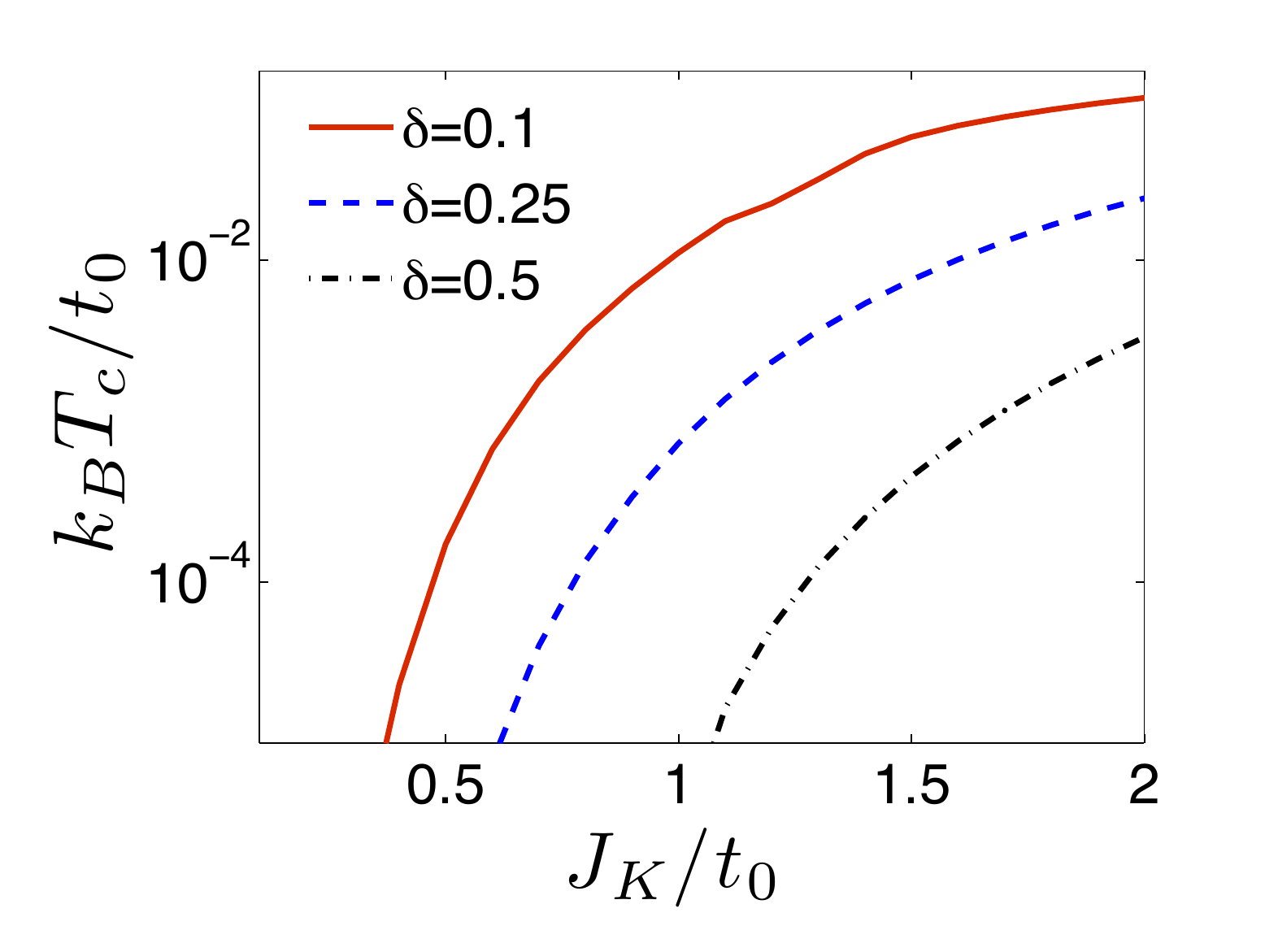}
\vspace{-0.3cm}
\caption{(color online) The transition temperature $k_B T_c/t_0$ in the p-wave
 triplet channel as a function of the Kitaev interaction strength $J_K/t_0$
 for different doping levels $\delta=0.1, 0.25, 0.5$, at $J_H=0$ \cite{fluctuation}.} 
\label{Tcvsdelta}
\end{figure}

The mean field Hamiltonian is obtained by replacing the spin-singlet and
spin-triplet operators by their expectation values in the usual fashion. We
denote the spin-singlet and spin-triplet expectation values by
$\Delta_{\gamma}$ and $d_{\gamma}^\rho$ respectively, where $\gamma$
corresponds to the bond direction, and $\rho$ the component of the mean-field
triplet vector, such that 
\begin{eqnarray}
\Delta_{\gamma} &=& -\frac{1}{\sqrt{2}}\bigl(J_H-\frac{J_K}{4}\bigr)\langle s_{ij}\rangle \ \ \  \textrm{for} \  \   \gamma \ \textrm{bonds}, \nonumber\\
d_{\gamma}^\rho &=& \frac{J_K}{4\sqrt{2}}\langle t_{ij,\rho}\rangle  \ \ \
\textrm{for} \  \   \gamma \ \textrm{bonds}. 
\end{eqnarray}
A different decoupling of the Kitaev interaction into both hopping and pairing channels was used in \cite{BuNa11} to describe the undoped Kitaev model. Here we consider only the pairing channel, which is justified in the limit of reasonably large doping. 

At the critical temperature for the superconducting phase transition the order
parameter vanishes and we obtain linearized gap equations \cite{Sigrist}. The
spin singlet case was previously solved in 
Ref.~\onlinecite{BlSc07}, obtaining a mixed superconducting phase of d-wave
intraband 
pairing and p-wave interband pairing. Following a similar procedure we obtain
 linearized gap equations also for the spin-triplet order parameters. We
find that the  $d^\rho_\gamma$ with different $\rho$ are not coupled in these
equations 
and therefore we can conveniently write the linearized gap equations for
triplet order parameters using three stability matrices $M_{x, y, z}$. The
critical temperatures for the different channels can be obtained by finding
the largest temperature where at least one of the eigenvalues of the stability
matrix is equal to 1. The possible solutions of the order parameter near the
critical temperature are linear combinations of the eigenvectors corresponding
to this eigenvalue. 

We find that the stability matrix for $d^x_\gamma$ is
\begin{equation}
M_x = \frac{J_K}{4}
\begin{pmatrix}
-B & C & C\\
-C & B & C\\
-C & C & B
\end{pmatrix}.
\end{equation}
The stability matrices $M_y$ and $M_z$ can be obtained from $M_x$ by cyclically changing the column of negative signs. The matrix elements are $B = A_{i=j}$ and $C = A_{i\neq j}$, where

\begin{figure}[tbp]
\centering\includegraphics[width=7.1cm]{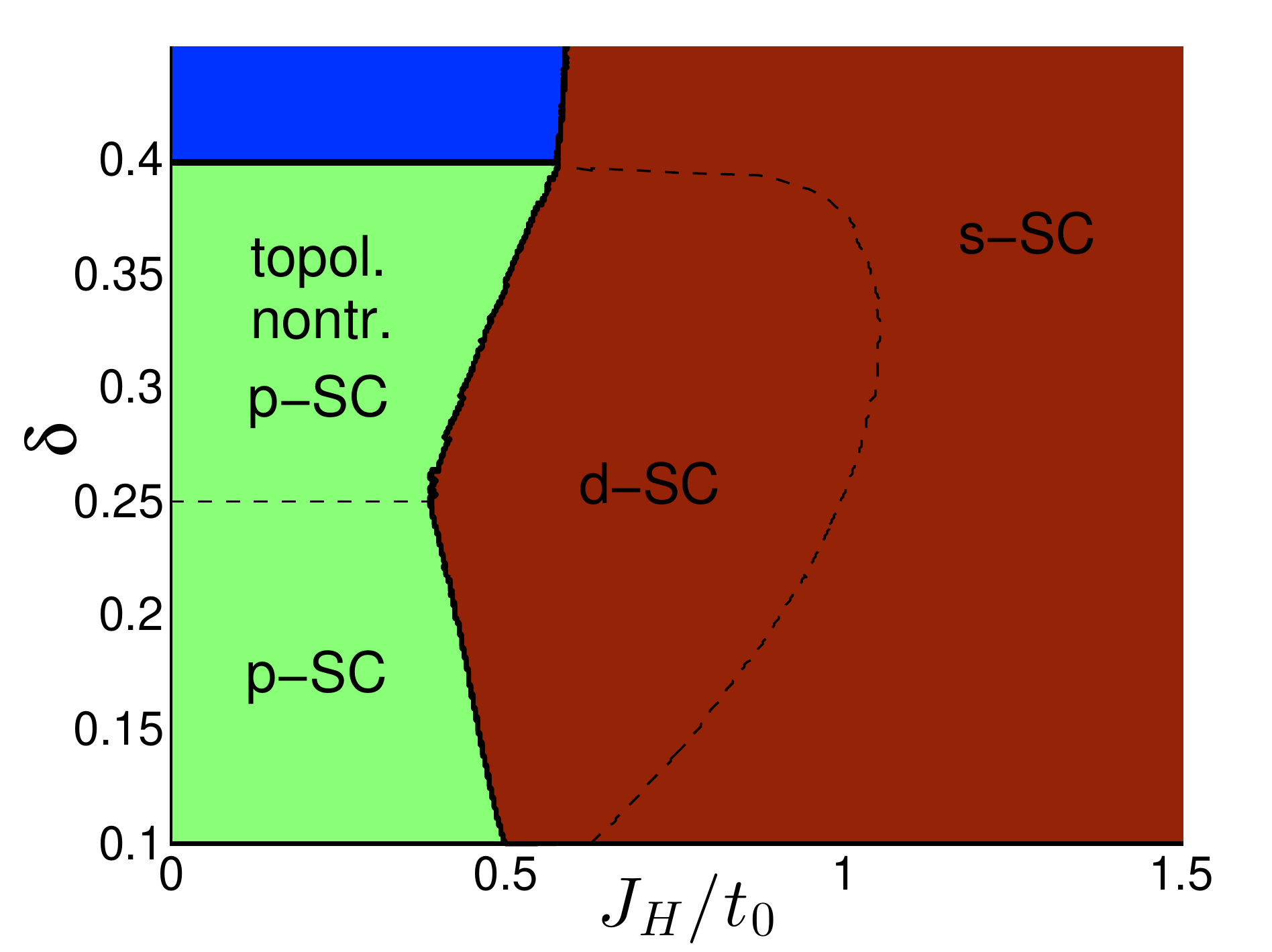}
\caption{(color online) Phase diagram  for the condensation in the
 spin-singlet (red/dark gray regions marked as s-SC and d-SC) and spin-triplet (green/light gray regions) channels as a function of
 Heisenberg interaction $J_H$ and doping $\delta$, for Kitaev interaction 
strength
 $J_K= t_0$.  In the singlet channel both d-wave and s-wave order parameters can exist depending on the doping and interaction strength. The
 d-wave order parameter breaks the time-reversal symmetry. Within the p-wave 
triplet phase, there is a  
 topological phase transition (indicated by dashed line) between
 topologically trivial $\delta < 0.25$ 
 and topologically non-trivial $\delta>0.25$ time-reversal invariant
 superconducting states. The (blue/dark gray) region in the upper left corner of the phase-diagram denotes the parameter regime where $k_B
 T_c <10^{-4} t_0$.} 
\label{phasediagram}
\end{figure}

\begin{equation}
\begin{split}
A_{ij} &= \frac{1}{2N}\sum_{\vec{q}} \biggl[\biggl(\frac{\tanh(\beta_c\xi_1/2)}{2\xi_1} + \frac{\tanh(\beta_c\xi_2/2)}{2\xi_2}\biggr)\\
&\ \ \ \ \ \ \ \ \ \ \ \ \ \ \ \cdot\sin(\vec{\delta}_i\cdot\vec{q}-\phi)\sin(\vec{\delta}_j\cdot\vec{q} - \phi) \\
&+\frac{\sinh(\beta_c\mu)\cos(\vec{\delta}_i\cdot\vec{q} - \phi)\cos(\vec{\delta}_j\cdot\vec{q} - \phi)}{2\mu\cosh(\beta_c\xi_1/2)\cosh(\beta_c\xi_2/2)}\biggr],
\end{split}
\end{equation}
in which $\beta_c=1/(k_BT_c)$, $\vec{\delta}_i$ denote the nearest neighbor
vectors, $\xi_{1(2)} = \pm|t(\vec{q})|-\mu$ are the single particle
dispersions,  $t(\vec{q})=t \sum_j e^{i \vec{\delta}_j \cdot \vec{q}}$  and
$\phi = \arg[t(\vec{q})]$. 

Because of the symmetry relation of the matrices $M_{x, y, z}$, the eigenvectors of the linearized gap equation with largest critical temperature are three-fold degenerate.  By denoting the solutions as $\mathbf{d}=(d^x_x, d^x_y, d^x_z, d^y_x, d^y_y, d^y_z, d^z_x, d^z_y, d^z_z)$, the linearly independent solutions are
\begin{eqnarray} 
\mathbf{d}_1=(0, -1, 1, 0, 0, 0,0, 0, 0), \nonumber\\
\mathbf{d}_2=(0, 0, 0, 1, 0, -1,0, 0, 0) , \nonumber\\
\mathbf{d}_3=(0, 0, 0, 0, 0, 0,-1, 1, 0).
\end{eqnarray}

We now compare the critical temperatures for superconducting condensation
in the spin-singlet and spin-triplet channels. In Fig.~\ref{Tcvsdelta} we show
the critical temperatures for the onset of spin-triplet superconductivity at
various doping levels. In the spin-singlet channel, the Kitaev interaction 
affects the pairing only by renormalizing the interaction strength $J_H$ and therefore the 
critical temperatures for singlet-superconductivity can be obtained using 
the theory of 
Ref.~\onlinecite{BlSc07} that predicts a 
time-reversal violating d-wave state (analogous to that found for a doped 
Heisenberg model in a triangular lattice \cite{Bas03,Wan04,Khaliullin})  or s-wave state depending on the interaction strength and doping. 
The phase diagram is shown in Fig.~\ref{phasediagram}, where the dominant
condensation channel is computed  as a function of doping and Heisenberg
interaction $J_H$ for fixed Kitaev interaction strength $J_K=
t_0$. Interestingly, we find that the phase transition between  triplet and
singlet superconductivity appears at the relative interaction strength
$J_H/J_K\simeq1/2$. We note that in the absence of doping the topologically 
interesting Kitaev spin liquid phase exists only in the range 
$J_H<J_K/8$ \cite{Cha+10,Reu+11}. This indicates  that the spin-triplet 
superconductivity is very robust to the presence of the 
Heisenberg 
interaction, and therefore  might be observable in doped iridates.

\begin{figure}[tbp]
\centering\includegraphics[width=7.2cm]{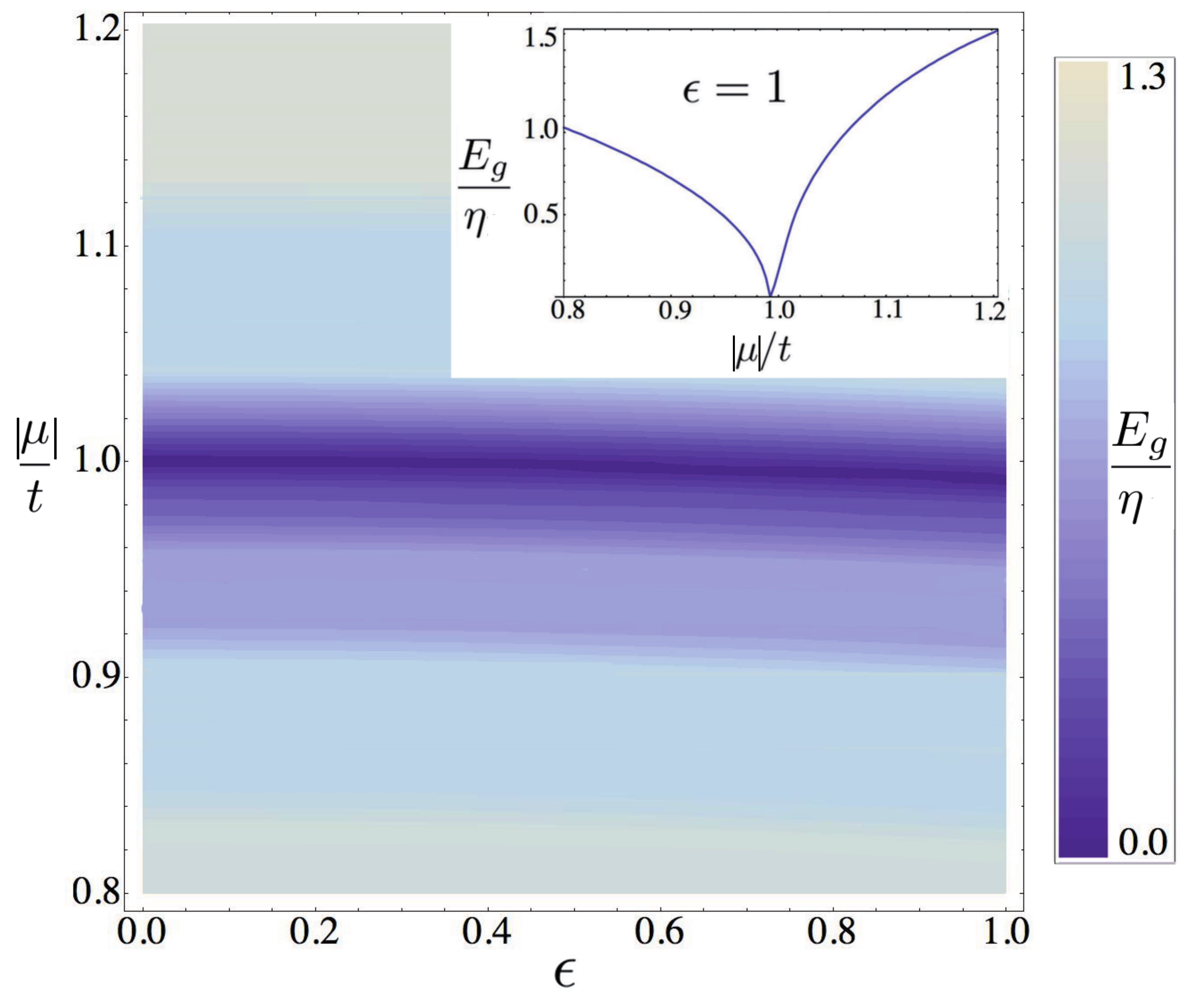}
\caption{(color online) The minimum energy required to excite quasiparticles
 as a function of the dimensionless interband order parameter strength
 $\epsilon$ and chemical potential $|\mu|$ for $\eta=0.042 t$. The
 interband order parameter is adiabatically switched on when $\epsilon$
 changes from  $\epsilon=0$ to $\epsilon=1$. The inset demonstrates that for
 each value of $\epsilon$ the energy gap closes in the vicinity of $|\mu|=t$
 indicating a transition between the topologically trivial and nontrivial
 states.} 
\label{robustness.fig}
\end{figure}
\begin{figure}[tbp]
\centering\includegraphics[width=6.5cm]{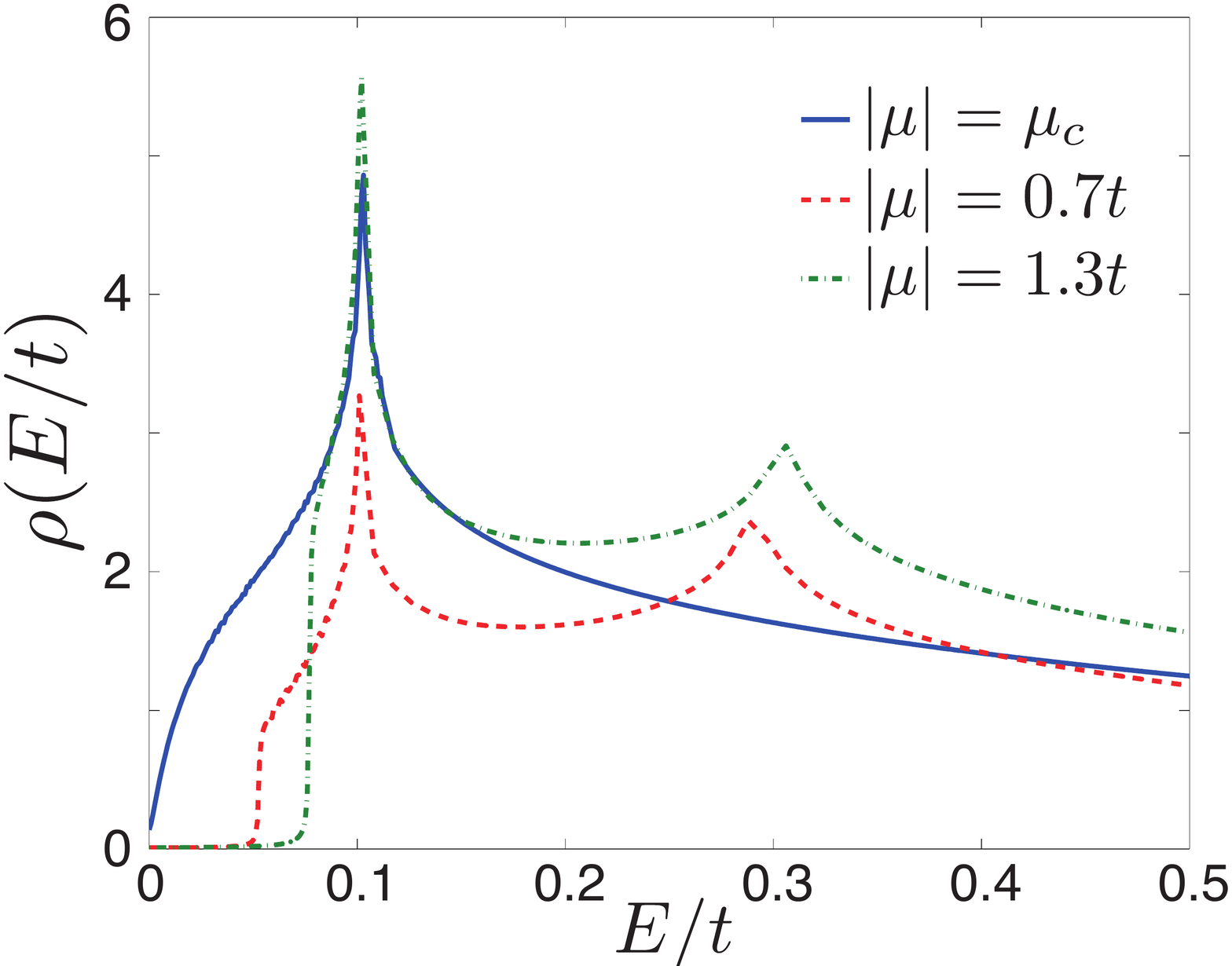}
\caption{ (color online) Integrated density of states for $\eta=0.042 t$ and
 chemical potentials $0.7 t$, $\mu_c$, and $1.3 t$. The critical value of the
 chemical potential is $\mu_c = 0.9932t.$} 
\label{Dos}
\end{figure}

The order parameter for triplet superconductivity at temperatures below the
$T_c$ can now be calculated from the nonlinear gap
equations. We assume that the order parameter is a linear combination of the
three degenerate solutions to the linearized gap equations, and so we can
write $\dvec = (\eta_1\dvec_1+\eta_2\dvec_2+\eta_3\dvec_3)$.  
Solving the gap equations iteratively from different initial conditions, we
find four solutions  $(\eta_1,\eta_2,\eta_3) = \eta(1,\pm1, \pm
i)$. These solutions  have large basins of attraction. Moreover,  
the real parts of the eigenvalues of the stability matrices obtained by
linearizing the gap equation around these solutions are smaller than one,
indicating stability.  Within our mean field approach, the magnitude of $\eta$ depends on  temperature, doping, and 
the interaction
strength $J_K$ \cite{fluctuation}. For $k_B T=10^{-4} t$, $|\mu|=1.2 t$ (corresponding to
$\delta \approx 0.38$) and $J_K=2.3 t_0$, we obtain   $\eta \approx 0.042
t$. This value of $\eta$ is used for Figs.~\ref{robustness.fig} and \ref{Dos}.

By expressing the intra- and interband order parameters using the convention $ i (\dvec(\vec{q}) \cdot\vec{\sigma}) \sigma_2$ \cite{Sigrist}, we obtain expressions for the intra- and interband $\dvec$-vectors in momentum space.
Expanding these solutions around the $\Gamma$ point, we obtain 

\begin{equation}
\begin{split}
&\dvec_{11} (\vec{q})= i\eta \biggl(-\frac{q_x}{2}  + \frac{\sqrt{3} q_y}{2}, \pm(\frac{q_x}{2}+\frac{\sqrt{3}q_y}{2}), \pm q_x\biggr),\\
&\dvec_{12}(\vec{q}) = \frac{ \eta}{24}\biggl(-3q_x^2 + 2\sqrt{3}q_xq_y+3q_y^2,\\
&\qquad \pm(-3q_x^2-2\sqrt{3}q_xq_y+3q_y^2), \mp 4\sqrt{3}q_xq_y\biggr),
\end{split}
\label{dvec.eq}
\end{equation}
where $\dvec_{22}=-\dvec_{11}$ are the intraband pairing vectors, and
$\dvec_{21} = -\dvec_{12}$ are the interband pairing vectors. 
Using these $\dvec_{ij}(\vec{q})$-vectors we find that the numerical solution
of the nonlinear gap equations is that of a {\it time-reversal symmetric odd 
parity} phase in the dominant intraband channel, and a 
{\it time-reversal symmetric even parity} phase in the interband channel. Both
order parameters retain the 
six-fold rotational symmetry of the lattice, and so the phases obtained obey
all the underlying symmetries of the microscopic model \cite{You11}.   

The order parameter $\dvec_{11}$ in Eq.~(\ref{dvec.eq})  lies on a circle with
fixed $|\dvec_{11}|$ when $\vec{q}$ rotates around a circle in the 
$(q_x,q_y)$--plane.  
The axis $\rvec_d$ around which $\dvec_{11}$ rotates depends on the choice of signs in Eq.~(\ref{dvec.eq}). For example, the 
choice $(-,+)$ corresponds to a rotation axis along the $(1,1,1)$ direction. 
By virtue of a global transformation of the 
spin quantization axis in the model Eq.~(\ref{hkh}), the $\dvec_{11}$ vector
is rotated into the $xy$--plane,  
and describes a $q_x - i q_y$ pairing for spin-up and a $q_x 
+ i q_y$ pairing for spin-down (equivalent to the 
B-phase of superfluid $^3$He) components of the order parameter. 
In the original model Eq.~(\ref{hkh}), this corresponds to $q_x - i q_y$ pairing
for spins pointing in the $(1,1,1)$ direction and $q_x  + i q_y$ pairing for
spins pointing in the $(-1,-1,-1)$ direction, see Fig.~\ref{pairing.fig}(a).

In the absence of interband coupling and in the weak pairing limit, the value of the $Z_2$ invariant of a
time-reversal invariant topological superconductor is determined by the
topology of the Fermi surface  
\cite{Sato09a,Sato09,Qi2010,YTada,Sato10}. More precisely, its value is
determined by the parity of the number of  
time-reversal invariant points below the Fermi level. On the honeycomb
lattice, the Fermi surface  
topology changes from hole pockets around the two inequivalent $K$  points to
a particle-like Fermi surface around the $\Gamma$ point when 
the chemical potential is tuned through the critical value $|\mu|=\mu_c = t$
(corresponding to the critical value of hole doping $\delta_c=0.25$).  
For $|\mu| < \mu_c$, there are four time-reversal invariant points below the
Fermi level (the three $M$ points 
and the $\Gamma$ point shown in Fig.~\ref{pairing.fig}), and the
superconductor is in a topologically trivial phase.  
At the special point $|\mu| =\mu_c $ in the single particle dispersion, the
Fermi surface touches the three inequivalent time-reversal symmetric $M$
points.  
As a consequence, for $|\mu| > \mu_c$, the only time-reversal invariant point
enclosed by the Fermi surface is the $\Gamma$ point. 
Because the  intraband order parameter has odd parity, the superconductor in
the absence of interband order parameter is guaranteed to be a topologically
non-trivial superconductor \cite{Sato09a,Sato09,Qi2010,YTada,Sato10}, in the
class DIII \cite{Schny+08}. Adiabatically switching on the interband pairing,
we find that the 
quasiparticle dispersion indeed remains gapped in each topological regime and
that  the value of the critical chemical potential $\mu_c$, where the
quasiparticle dispersion becomes gapless,  is renormalized downwards.
Therefore, the topological phase transition  
remains robust in the presence of the even parity interband phase (see
Fig.~\ref{robustness.fig}). This phase transition is shown with a dashed line
in Fig.~\ref{phasediagram}. 

The topological phase transition can be  observed by measuring
the density of states as a function of energy, e.g., by NMR and tunneling experiments. Fig.~\ref{Dos} shows the
density of states as a function of energy for three different values of
chemical potential. Above and below $|\mu|=\mu_c$ the density of states vanishes
at low energies resulting in an energy gap, which is of the order of
$\eta$. On the other hand, at the phase transition  $|\mu|=\mu_c$  the energy
gap closes resulting in an increased density of states at low energies.

A few remarks are in order concerning the relevance of our results for the actual materials. First, we have considered here the nearest-neighbor hopping Hamiltonian, but our results are in fact robust with respect to changes in the band structure. We have verified that all our qualitative results remain valid if a second nearest neighbor hopping $t'=0.1t$ is introduced. Second, the necessary values of the Kitaev interaction strength for the existence of topological p-wave superconductivity might seem quite large $J_K>0.7 t_0$. Nevertheless, these values are not unrealistic for the iridates, because the relation between the exchange interaction strength and the effective nearest-neighbor hopping in iridates is not as simple as, for example, in cuprates. In systems with strong spin-orbit coupling, the electron collects a phase factor during the superexchange process giving rise to a Kitaev interaction, and, at the same time, the effective nearest-neighbor hopping amplitude is strongly reduced by negative quantum interference. We also notice that approximately similar values of the exchange interaction strength have been discussed for graphene  \cite{BlSc07}.

In summary, we have studied the competition of spin-singlet d-wave
superconductivity and spin-triplet p-wave superconductivity in the hole-doped
Kitaev-Heisenberg 
model on a honeycomb lattice. We have shown
that the Kitaev interaction favors a time-reversal invariant $Z_2$ topological
superconductivity, which is robust to the presence of an isotropic Heisenberg
interaction $J_H<J_K/2$. Because this topological phase supporting pairs of
counter-propagating Majorana modes persists over a much wider region of the 
phase diagram than the Kitaev spin liquid phase itself in the absence of
doping, it might be easier to find materials where it becomes observable.

We would like to thank A.~Vishwanath and A.~Schnyder for useful communications and comments.
  Financial support by the BMBF is acknowledged.


\begin{thebibliography}{99}
\bibitem{fluctuation} Within the present mean field 
approach, $J_H$ below its critical value does not affect $T_c$ and the amplitude of the order parameter in the triplet channel, but fluctuation corrections from $J_H$ may 
reduce them.
\end{thebibliography}
\end{document}